\documentclass[aps,prb,twocolumn,groupedaddress,showpacs]{revtex4}

\usepackage{times}
\usepackage{graphicx}
\usepackage{amsfonts}
\usepackage{amsmath, amsthm, amssymb}
 
\newcommand{\ve}[1]{\mathbf{#1}}
\newcommand{\vk}{\ve{k}} 

\newcommand{\e}[1]{\mathrm{e}^{#1}}

\newcommand{\ie}{\textit{i.e. }}
\newcommand{\eg}{\textit{e.g. }}
\newcommand{\etal}{\emph{et al.}}
\def\i{\mathrm{i}}

\begin{document}
\title[Conductance spectra of ferromagnetic superconductors: Quantum
    transport in a ferromagnetic metal/non-unitary ferromagnetic
    superconductor junction]{Conductance spectra of ferromagnetic superconductors: Quantum
    transport in a ferromagnetic metal/non-unitary ferromagnetic
    superconductor junction}
\author{J. Linder}
\affiliation{Department of Physics, Norwegian University of
Science and Technology, N-7491 Trondheim, Norway}
\author{M. S. Gr{\o}nsleth}
\affiliation{Department of Physics, Norwegian University of
Science and Technology, N-7491 Trondheim, Norway}
\author{A. Sudb{\o}}
\affiliation{Department of Physics, Norwegian University of
Science and Technology, N-7491 Trondheim, Norway}
\affiliation{Centre for Advanced Study, Norwegian Academy of Science and papers, Drammensveien 78, N-0271 Oslo, Norway.  }\date{Received \today}

\begin{abstract}
Recent findings of superconductors that 
simultaneously exhibit multiple spontaneously broken 
symmetries, such as ferromagnetic order or lack of an 
inversion center and even 
combinations of such broken symmetries, 
have led to much theoretical and experimental research. 
We consider quantum transport in a junction consisting of a 
ferromagnetic metal and a non-unitary ferromagnetic 
superconductor. It is shown that the conductance spectra 
provides detailed information about the superconducting gaps, 
and is thus helpful in determining the pairing symmetry 
of the Cooper pairs in ferromagnetic superconductor.
\end{abstract}
\maketitle
\section{Introduction}
Spontaneous symmetry breaking in condensed matter systems 
ranks among the most profound emergent phenomena in 
many-body physics. 
Multiple spontaneously broken symmetries are not only of interest in terms of studying properties of specific condensed matter systems, but also due to the fact that it may provide clues 
for what could be expected in other systems in vastly 
different areas of physics. A first attempt \cite{ginzburg} at describing 
the physics of magnetic spin-singlet superconductors was made 
long ago, and the discovery of ferromagnetic superconductors (FMSCs) \cite{saxena,aoki} displaying 
coexisting superconductivity (SC) and ferromagnetism (FM) 
well below the Curie temperature, has produced a realization 
of a physically rich system that exhibits simultaneously broken $O(3)$ and $U(1)$ symmetries. Spontaneous breaking of symmetry 
is responsible for a wide range of physical phenomena, such as 
the mass differences of elementary particles, phase transitions in condensed-matter systems, and even emergent phenomena in biology \cite{pwa}. In many cases, the phenomena can in fact be described by the same equations. Thus, a study of ferromagnetic superconductors is of interest not only in terms of having 
an obvious potential for leading to novel devices in for instance nanotechnology and spintronics, but also from a fundamental 
physics point of view.\\
\indent A successful model describing a FMSC demands that two important issues are adressed: \textit{i}) the symmetry of the pairing state, and \textit{ii}) whether the superconducting and ferromagnetic order parameters are coexistent or phase-separated  in space-time. Cooper pairs in conventional superconductors ($s$-wave) are spin-singlets. Thus, $s$-wave pairing and uniform FM are antagonists \cite{varma,shen}. Spin-triplet Cooper pairs, however, can carry a net magnetic moment. Such Cooper pairs are presently the prime candidate for explaining the coexistence of FM and SC 
in \eg UGe$_2$, and URhGe \cite{saxena,aoki}. In these materials, 
SC occurs deep within the ferromagnetic phase. It is therefore natural to view the SC pairing as originating with electrons 
that also contribute to FM.  Thus, the electrons responsible 
for FM below the Curie temperature $T_\text{M}$ condense 
into Cooper pairs with magnetic moments aligned along the magnetization below the critical temperature $T_\text{c}$. 
While spin-singlet superconductivity coexisting with uniform ferromagnetism appears to be unlikely, it could coexist with 
helically ordered magnets. Tunneling phenomena 
in such systems have indeed been considered theoretically \cite{kulic,eremin}. This is, however, physically quite 
different from the case we will study in this paper.  \\
\indent Bulk FMSCs are expected to display an unusual feature, namely 
the spontaneous formation of an Abrikosov vortex lattice \cite{tewari}. Uniform superconducting phases have 
also been investigated \cite{shopova}, but in a bulk system 
it seems reasonable to assume that this must be associated 
with a nonuniform magnetic state \cite{kulic,eremin}. A 
key variable determining whether a vortex lattice appears 
or not seems to be the strength of the internal magnetization
$\mathbf{m}$ \cite{mineev2}. The current experimental data \cite{aoki} available for URhGe apparently do not settle this 
issue unambiguously, while uniform coexistence of FM and SC 
appears to have been experimentally verified \cite{kotegawa} 
in UGe$_2$. Moreover, a bulk Meissner state in the FMSC RuSr$_2$GdCu$_2$O$_8$ has been reported \cite{bernhard}. 
No consensus has so far been reached concerning the correct 
pairing symmetry for the FMSCs, although theoretical 
considerations strongly suggest that a non-unitary state 
is favored \cite{huxley,samokhin,machida}. In terms of the $\mathbf{d}_\vk$-vector formalism (see below), this means that the relation $\i(\mathbf{d}_\vk\times\mathbf{d}_\vk^*)\neq0$ is satisfied, which is equivalent to saying that the Cooper pairs carry a net spin \cite{leggett1975}. The study of pairing symmetries in unconventional superconductors has a long 
tradition \cite{pair} where tunneling currents have played 
a crucial role. For the case of spin-triplet non-magnetic superconductors, theoretical studies of tunneling to a normal 
metal or ferromagnet have been suggestive in terms of 
establishing the correct pairing symmetry for the 
superconductor \cite{bolech, stefanakis}.\\
\indent In this paper, we consider quantum transport between two thin films of a non-unitary FMSC and an easy-axis ferromagnet, respectively. We demonstrate how 
the resulting conductance spectra can be exploited to obtain useful information about the superconducting gaps. The SC and FM order parameters are assumed not to be phase-separated.
Moreover, the choice of a thin film FMSC is motivated by the 
fact that the pair-breaking orbital effect on Cooper pairs 
with an in-plane magnetization is suppressed, and that one 
will avoid vortices present in the compound if the thickness 
$t$ of the film is smaller than the diameter of a vortex \cite{meservey}, $t<\xi_0\ll\lambda_0$. Here, $\xi_0$ is the coherence length (typical size of Cooper pairs) while $\lambda_0$ is the penetration depth (typical radius of vortex). In the cases \cite{saxena, aoki} of UGe$_2$ and URhGe, this amounts to $t\sim10$ nm which is well within reach 
of current experimental techniques. \\
\indent We organize our work as follows. In Sec. \ref{sec:model}, we establish the model and formalism which we will apply to the problem. Results are given in Sec. \ref{sec:results}, in addition to a discussion of our findings. Concluding remarks are given in Sec. \ref{sec:conclusion}
\section{Model and formulation}\label{sec:model}
\begin{figure}[h!]
\centering
\resizebox{0.48\textwidth}{!}{
\includegraphics{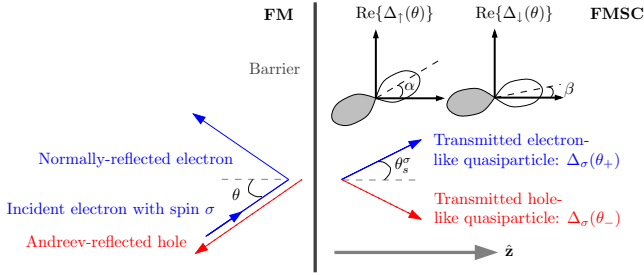}}
\caption{(Color online). Model system of a FM-FMSC interface in a slab-geometry
in the clean limit. Retroreflection symmetry is not broken, since the hole carries the same spin as the incident electron. We have sketched gaps corresponding to the analog \cite{leggett1975} of the $A2$-phase in 
liquid $^3$He.}
\label{fig:model}
\end{figure}

Our model is illustrated in Fig. \ref{fig:model}, where the 
superconducting pairing symmetry is equivalent to that 
of an $A2$-phase in $^3$He [see Ref.~\onlinecite{leggett1975} 
and the discussion below Eq. (\ref{eq:bdg})] .
\par
The Bogoliubov-de Gennes (BdG) equations for a FMSC \cite{bj1} 
may be written in compact matrix form
\begin{equation}\label{eq:bdg}
\begin{pmatrix}
\hat{\mathcal{M}}_\vk & \hat{\Delta}_\vk \\
\hat{\Delta}^*_\vk & -\hat{\mathcal{M}}_\vk
\end{pmatrix} 
\begin{pmatrix}
u_{\vk\sigma} \\
v_{\vk\sigma}
\end{pmatrix}
= E_{\vk\sigma}
\begin{pmatrix}
u_{\vk\sigma} \\
v_{\vk\sigma}
\end{pmatrix},
\end{equation}
where we have introduced the 2$\times$2 matrices
\begin{align}
\hat{\mathcal{M}}_\vk &= \varepsilon_\vk\hat{1} - \hat{\sigma}_z U_\mathrm{R},\notag\\
\hat{\Delta}_\vk &= (\Delta_0\hat{1} + \hat{\boldsymbol{\sigma}}\cdot\mathbf{d}_\vk)\i\hat{\sigma}_y,
\end{align} 
in addition to $u_{\vk\sigma} = [u_{\vk\uparrow\sigma}\; u_{\vk\downarrow\sigma}]^\mathrm{T}$, $v_{\vk\sigma} = [v_{\vk\uparrow\sigma}\; v_{\vk\downarrow\sigma}]^\mathrm{T}$. Here, $\varepsilon_\vk$ is a kinetic energy term, $\hat{\boldsymbol{\sigma}}=(\hat{\sigma}_x,\hat{\sigma}_y,\hat{\sigma}_z)$ is a vector containing the Pauli 
matrices, $U_\mathrm{R(L)}$ describes the magnetic 
exchange energy of right (left) part of the system. 
Moreover, the $\mathbf{d}_\vk$ vectors are given by 
\begin{equation}
\mathbf{d}_\vk =  \frac{1}{2}[\Delta_{\vk\downarrow\downarrow}-\Delta_{\vk\uparrow\uparrow},-\i(\Delta_{\vk\downarrow\downarrow} + \Delta_{\vk\uparrow\uparrow}),2\Delta_{\vk\uparrow\downarrow}],
\end{equation}
and  $\Delta_0$ and $\Delta_{\vk\alpha\beta}$ are the superconducting spin-singlet and spin-triplet order 
parameters, respectively. Finally, $u_{\vk\sigma}$ 
and $v_{\vk\sigma}$ are the wave-function solutions 
with eigenvalue $E_{\vk\sigma}$. In the following, 
we set the $\vk$-independent singlet amplitude $\Delta_0$ 
to zero, as we do not consider the case of coexistent 
$s$-wave SC and FM \cite{machida}. Also, 
the opposite-spin triplet pairing giving rise to the gap $\Delta_{\vk\uparrow\downarrow}$ is in general believed to be suppressed, since it requires interband pairing of 
Zeeman-split electrons \cite{aoki}. Hence, our model is 
that of a non-unitary spin-triplet state with equal-spin 
pairing, \ie $\Delta_{\vk\uparrow\downarrow} = 0, \Delta_{\vk\sigma\sigma} \neq 0$, equivalent to the 
$A2$-phase in liquid $^3$He (see \eg Ref.~\onlinecite{leggett1975}) with a non-vanishing magnetic moment associated with the Cooper pairs.\\
\indent As indicated in Fig. \ref{fig:model}, the reflected and 
transmitted electron- and hole-like excitations feel 
different pairing potentials due to the orbital symmetry 
of the superconducting gaps. The angle into which they are
scattered depends on the spin $\sigma$ of the incident 
electron, since there is a magnetic exchange energy 
$U_\mathrm{R}$ present in the superconducting state. In 
the following, we will consider the momentum of the 
quasiparticles as fixed on the Fermi surface for spin 
$\sigma$ so that the superconducting gaps correspondingly 
only depend on the direction of momentum, \ie $\Delta_{\vk\sigma\sigma}\to\Delta_\sigma(\theta_s^\sigma)$ 
where $\theta_s^\sigma$ is defined in Fig. \ref{fig:model}. Throughout this paper, we shall insert the superconducting gap 
\textit{a priori} instead of solving it self-consistently, 
in order to obtain analytical formulas. This is justified by the fact that detailed calculations taking into account the modification of the pair potential near the barrier demonstrate that no new qualitative features appear in the conductance spectra compared to when the gap is modelled by a step function at the interface \cite{suppression}. 
However, the proximity effect at a FM/SC interface may cause two important phenomena to occur: \textit{i)} induction of an SC order parameter (possible mixture of singlet and triplet) in the FM region \cite{eschrig} and \textit{ii)} the formation of midgap surface states on the interface \cite{hu}, leading to a suppression of the OP in the vicinity of the interface. The competition and coexistence of these two phenomena has been studed in Ref.~\onlinecite{tanaka2}. In this work, we will mainly be concerned with a SC pairing symmetry analogous to the A2-phase in liquid $^3$He, for which the latter of these effects is only present for a specific trajectory of the incoming electrons. Thus, we believe our results capture the essential qualitative features of the conductance spectra, although a self-consistent approach would be required in order to obtain the entire picture.\\
\indent For the simplest model that illustrates the physics, we have  chosen a two-dimensional FM-FMSC junction with a barrier 
modelled by the potential $V(\mathbf{r}) = V_0\delta(z)$ and superconducting gaps $\Delta_\sigma(\theta_s^\sigma,\mathbf{r}) = \Delta_\sigma(\theta_s^\sigma)\Theta(z)$. Here, $\delta(z)$ and $\Theta(z)$ represent the Kronecker delta- and Heaviside 
functions, respectively. Solving the BdG-equations and applying 
the BTK formalism \cite{btk}, one finds that our system in Fig. \ref{fig:model} is described by the wave-functions for particles and holes with spin $\sigma$ in the ferromagnet $(\psi^\sigma)$ 
and FMSC $(\Psi^\sigma)$ side of the barrier. Explicitly, the total wave-function $\Psi_\text{tot}^\sigma(z) = \Theta(-z)\psi^\sigma(z) + \Theta(z)\Psi^\sigma(z)$ then has the components
\begin{widetext}
\begin{align}\label{eq:wavefunction}
\psi^\sigma(z)   &= \e{\i k^\sigma\sin\theta y}\Big[ \begin{pmatrix}
1\\
0
\end{pmatrix}\e{\i k^\sigma\cos\theta z} + r_e^\sigma(E,\theta)\begin{pmatrix}
1\\
0
\end{pmatrix}
\e{-\i k^\sigma\cos\theta z} + r_h^\sigma(E,\theta)\begin{pmatrix}
0\\
1
\end{pmatrix}\e{\i k^\sigma\cos\theta z}\Big],\notag\\
\Psi^\sigma(z) &=  \e{\i q^\sigma\sin\theta y}\Big[t_e^\sigma(E,\theta)\begin{pmatrix}
u_\sigma(\theta_{s+}^\sigma)\\
v_\sigma(\theta_{s+}^\sigma)\gamma^*_\sigma(\theta_{s+}^\sigma)
\end{pmatrix}
\e{\i q^\sigma\cos\theta_{s}^\sigma z} 
+ t_h^\sigma(E,\theta)\begin{pmatrix}
v_\sigma(\theta_{s-}^\sigma)\gamma_\sigma(\theta_{s-}^\sigma) \\
u_\sigma(\theta_{s-}^\sigma)
\end{pmatrix}\e{-\i q^\sigma\cos\theta_{s}^\sigma z}\Big],
\end{align}
\end{widetext}
with $\gamma_\sigma(\theta) = \Delta_\sigma(\theta)/|\Delta_\sigma(\theta)|$, $\theta_{s+}^\sigma = \theta_s^\sigma$, $\theta_{s-}^\sigma = \pi-\theta_s^\sigma$. The wave-vectors read $k^\sigma = [2m(E_\mathrm{F} + \sigma U_\mathrm{L})]^{1/2}$, $q^\sigma = [2m(E_\mathrm{F} + \sigma U_\mathrm{R})]^{1/2}$ 
while the spin-generalized coherence factors are
\begin{equation}
u_\sigma(\theta_{s\pm}^\sigma) [v_\sigma(\theta_{s\pm}^\sigma)]=\frac{1}{4}\{1+[-]\sqrt{1 - (|\Delta_\sigma(\theta_{s\pm}^\sigma)|/E)^2}\}^{1/2}.
\end{equation}
In writing down Eq. (\ref{eq:wavefunction}), we have implicitly incorporated conservation of group velocity and conservation of momentum parallell to the barrier, \ie $k^\sigma\sin\theta = q^\sigma\sin\theta_s^\sigma$. 

\section{Results and discussion}\label{sec:results}
The normal- and Andreev-reflection coefficients can be calculated by making use of the boundary conditions
\begin{align}
\textit{i)}\;& \psi^\sigma(0) = \Psi^\sigma(0),\notag\\
\textit{ii})\;& 2mV_0\psi^\sigma(0) = \mathrm{\partial}/\mathrm{\partial}z[\Psi^\sigma(z) - \psi^\sigma(z)]|_{z=0}. 
\end{align}
Let us introduce $Z=2mV_0/k_\mathrm{F}$ and 
\begin{equation}
\Upsilon_\pm^\sigma=q^\sigma\cos\theta_{s}^\sigma\pm k^\sigma\cos\theta \pm \i k_FZ,
\end{equation}
while $P_\sigma^\mathrm{L(R)} = (E_F +\sigma U_\mathrm{L(R)})/2E_F$ denotes the spin polarization on the left (right) side of the junction. Our calculations then lead to the explicit expressions for the Andreev- and normal-reflection coefficients for this FM-FMSC junction, namely
\begin{align}
r_e^\sigma &= -1 + 2k^\sigma\cos\theta[u_\sigma(\theta_{s+}^\sigma)u_\sigma(\theta_{s-}^\sigma)(\Upsilon_+^\sigma)^*\notag\\
&\;\;+v_\sigma(\theta_{s-}^\sigma)v_\sigma(\theta_{s+}^\sigma)\gamma_\sigma(\theta_{s-}^\sigma)\gamma_\sigma^*(\theta_{s+}^\sigma)(\Upsilon_-^\sigma)^* ]/D^\sigma, \notag\\
r_h^\sigma &= 4k^\sigma\cos\theta q^\sigma\cos\theta_s^\sigma v_\sigma(\theta_{s+}^\sigma)u_\sigma(\theta_{s-}^\sigma)\gamma_\sigma^*(\theta_{s+}^\sigma)/D^\sigma,
\end{align}
upon defining the quantity
\begin{align}
D^\sigma &= u_\sigma(\theta_{s+}^\sigma)u_\sigma(\theta_{s-}^\sigma)|\Upsilon_+^\sigma|^2 \notag\\
&\;\;- v_\sigma(\theta_{s-}^\sigma)v_\sigma(\theta_{s+}^\sigma) \gamma_\sigma(\theta_{s-}^\sigma)\gamma_\sigma^*(\theta_{s+}^\sigma)\times|\Upsilon_-^\sigma|^2.
\end{align}
In the limit of weak FM where all quasiparticle momenta equal $k_\mathrm{F}$, our results are found to be consistent with a spin-generalized augmentation of the equations in 
Ref. \onlinecite{tanaka}. In their general form, 
the above equations are novel results that include the effect 
of magnetism on the superconducting side of the barrier. 
As demanded by consistency, one obtains total reflection $|r_e^\sigma|^2 = 1$ when $\theta > \text{arcsin}(q^\sigma/k^\sigma)$. Having obtained the above quantities, one may calculate 
the conductance $G(E)$ of the setup (in units of the normal 
conductance, \ie $\Delta_\sigma(\theta) \to 0$). We find 
that it is given by 
\begin{equation}
G(E) = \sum_\sigma G^\sigma(E)/\sum_\sigma F^\sigma
\end{equation}
where we have defined the quantities
\begin{align}
G^\sigma(E)[F^\sigma] &= \int^{\pi/2}_{-\pi/2} \mathrm{d}\theta \cos\theta G^\sigma(E,\theta)[f^\sigma]P_\sigma^\mathrm{L} P_\sigma^\mathrm{R},\notag\\
G^\sigma(E,\theta) &= 1 + |r_h^\sigma(E,\theta)|^2 - |r_e^\sigma(E,\theta)|^2,
\end{align}
and $f^\sigma = 1 - |1 - 2k^\sigma\cos\theta/\Upsilon^\sigma_+|^2$.
We next demonstrate how the conductance spectra yields useful information about the superconducting gaps upon varying the relative orientation of the gaps, their magnitude, and the 
strength of the magnetic exchange energy on each side of the junction. To be specific, we first first consider the analog 
of the $A2$-phase in liquid $^3$He, such that the gaps may be written \cite{maeno} 
\begin{equation}
\Delta_\uparrow(\theta_{s\pm}^\sigma)  =  -\Delta_{\uparrow,0}\e{\i(\theta_{s\pm}^\sigma-\alpha)},\;
\Delta_\downarrow(\theta_{s\pm}^\sigma)  =  \Delta_{\downarrow,0}\e{-\i(\theta_{s\pm}^\sigma-\beta)}, 
\label{eq:gaps_1}
\end{equation}
as illustrated in Fig. \ref{fig:model}. We stress that $\theta_{s\pm}^\sigma$ is \textit{not} the global broken 
$U(1)$ superconducting phase, but rather an internal phase originating with the odd symmetry of the $p$-wave gaps 
(see Fig. \ref{fig:model}). The exchange energy of the 
FMSC will be kept fixed at $U_\mathrm{R} = 0.05E_\mathrm{F}$. 
In Fig. \ref{fig:fig1}, we have plotted the conductance spectra 
for a FM/FMSC junction for three distinct cases. We have 
defined the ratio between magnitude of the superconducting 
gaps as $R_\Delta = \Delta_{\uparrow,0}/\Delta_{\downarrow,0}$.
Introducing the dimensionless barrier strength $Z=2mV_0/k_\mathrm{F}$, where $k_\mathrm{F}$ is the Fermi 
momentum, we consider \textit{i)} the metallic contact case 
with no barrier ($Z=0$), \textit{ii)} the intermediate case 
with a moderate barrier ($Z=3$), and \textit{iii)} the 
tunneling limit with an insulator in the junction 
($Z\to\infty$). For each case \textit{i)-iii)}, we have 
allowed the magnetic exchange energy $U_\mathrm{L}$ to vary 
from weak, favoring $\uparrow$-spins ($U=0.05 E_\mathrm{F}$) 
to strong, favoring $\uparrow$-spins in one case ($U=0.5 E_\mathrm{F}$) and $\downarrow$-spins in the other case 
($U=-0.5 E_\mathrm{F}$). These are shown in descending order 
in each column. We have also considered the conductance 
$G(E)$ for several values of $\{\alpha,\beta\}$. For the
gaps chosen above, we  find that the $G(E)$ did not depend 
on different choices of these parameters. This can be 
understood by observing that the angular averaging 
over $G^\sigma(E,\theta)$ allows for factors 
$\e{-\i\alpha(\beta)}$ to be separated out on equal footing 
as the factor corresponding to the globally broken $U$(1) 
symmetry, whose value does not affect the conductance 
spectra. This will, however, not be the case for other 
possible triplet symmetries, as discussed below. \\
\indent An important, and obvious, feature of the 
conductance spectra is a characteristic behavior occuring 
at voltages corresponding 
to $E=\Delta_{\sigma,0}$, $\sigma=\uparrow,\downarrow$, 
where peaks are displayed. This offers the opportunity to 
utilise the conductance spectra to reap explicit information 
on the size of the superconducting gaps in the FMSC. From 
Fig. \ref{fig:fig1}, it is seen that the effect of increasing 
the exchange energy on the ferromagnetic side to $U_\mathrm{L} = \pm0.5E_\mathrm{F}$ is a sharpening of the peaks located at $E=\Delta_{\sigma,0}$, where $\sigma$ is the spin-species energetically favored by $U_\mathrm{L}$. Concomitantly, the 
peak located at $E=\Delta_{-\sigma,0}$ is suppressed. Such 
a response is consistent with what one would expect, since increased/decreased spin polarization on the ferromagnetic 
side enhances/suppresses the conductance of the corresponding 
spin component. In the tunneling limit (large $Z$), we see 
that the conductance spectra exhibits sharp transitions at $E=\Delta_{\sigma,0}$, corresponding to the sudden appearance 
of a tunneling current as the voltage exceeds the magnitude 
of the gaps. We have also considered $G(E)$ in the case 
of vanishing FM on the left side, \ie a N/FMSC junction. 
The results we find are very similar to the case of weak 
FM displayed in the upper row of Fig. \ref{fig:fig1}, and 
are therefore not displayed.

\begin{widetext}
\text{ }
\begin{figure}[h!]
\centering
\resizebox{1.0\textwidth}{!}{
\includegraphics{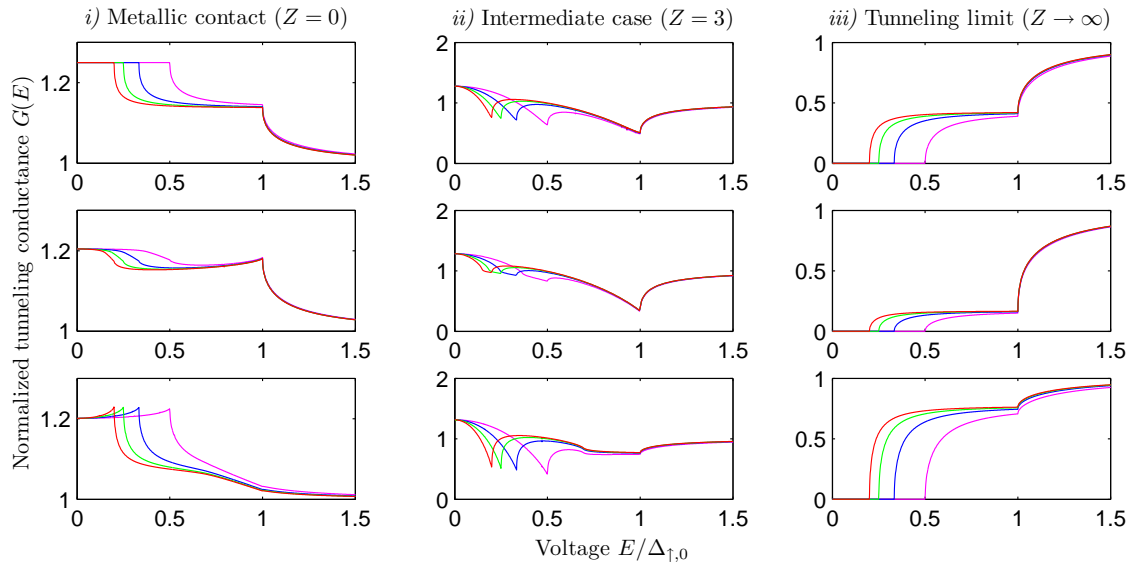}}
\caption{(Color online). Plot of the conductance $G(E)$ for 
a FM/FMSC junction in the case of a metallic contact, the 
tunneling limit, and an intermediate case. The gaps are given by
Eq. (\ref{eq:gaps_1}) in this case, for which $G(E)$ is found 
to be insensitive to $\{\alpha,\beta\}$. In the above, 
$\{\alpha,\beta\} = 0$. The columns for each case provide the spectrum for $U_\text{L} = 0.05E_\text{F}$, $U_\text{L} = 0.5E_\text{F}$, and $U_\text{L} = -0.5E_\text{F}$, in descending order. For each figure, we have plotted several ratios between the magnitude of the superconducting gaps. These are given by $R_\Delta=\{2,3,4,5\}$, represented by the magenta, blue, 
green, and red line, respectively.}
\label{fig:fig1}
\end{figure}

\end{widetext}

\begin{figure}[h!]
\centering
\resizebox{0.50\textwidth}{!}{
\includegraphics{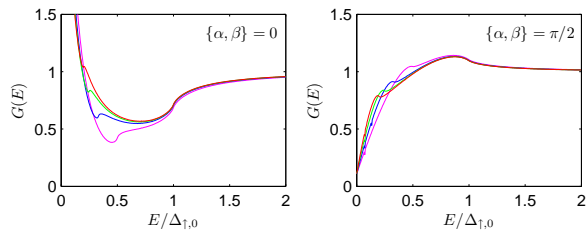}}
\caption{(Color online). Plot of the conductance $G(E)$ for FM/FMSC 
junction with gaps given by Eq. (\ref{eq:gaps_2}), 
for which $G(E)$ is sensitive to $\{\alpha,\beta\}$. 
We have chosen  $Z=3$, and $R_\Delta=\{2,3,4,5\}$ 
are represented by the magenta, blue, green, and 
red line, respectively.}
\label{fig:fig2}
\end{figure}

For the gaps in Eq. (\ref{eq:gaps_1}), the conductance 
was found to be insensitive to $\{\alpha,\beta\}$. However, 
in general this is different, and the orientation of the 
gaps relative to the barrier is essential in determining 
the conductance spectrum. We  illustrate this with a 
somewhat different choice of anisotropic gaps. When the superconducting gap is oriented in a fashion that leaves 
it invariant under inversion of the component of momentum perpendicular to the junction, $k_z\to-k_z$ in this case (equivalently $\theta\to\pi-\theta$), no zero-bias 
conductance peak (ZBCP) should be expected as there is no 
formation of current-carrying zero energy states \cite{hu}. However, when the gap changes sign under such an inversion 
of momentum, ZBCPs appear \cite{tanaka}. Since the momentum perpendicular to the junction of the hole-like excitation in 
the anisotropic superconductor is reversed compared to the electron-like excitation, a gap that satisfies 
$\Delta_\sigma(\theta) \neq \Delta_\sigma(\pi-\theta)$
will cause the hole to feel a different pairing potential than 
the electron-like excitation. This is the fundamental reason 
for the appearance of a ZBCP. However, in the present case of 
$p$-wave superconducting gaps analogous to the $A2$-phase in 
$^3$He, both gaps are \textit{asymmetric} under the operation $\theta\to\pi-\theta$ regardless of the value of $\{\alpha,\beta\}$ except for the single value $\theta=0$. Therefore, a small contribution to zero energy current-carrying states, \ie $G(0)\neq0$, will occur as shown in Fig. \ref{fig:fig1}.
However, this contribution does not lead to a ZBCP, for which 
all $\theta$ (see Fig. \ref{fig:fig1}) contribute to the 
formation of zero energy states due to the asymmetry of the 
gaps. In a model where the $p$-wave gaps are represented by 
the odd-symmetry analog of $d_{x^2-y^2}$-wave gaps, \ie 
\begin{equation}
\Delta_{\uparrow}(\theta) = \Delta_{\uparrow,0}\cos(\theta-\alpha),\; \Delta_{\downarrow}(\theta) = \Delta_{\downarrow,0}\cos(\theta-\beta), 
\label{eq:gaps_2}
\end{equation}
the formation of ZBCP will then be predictable from the 
orientation of the gaps as these can now display symmetry/antisymmetry/asymmetry 
when $\theta\to\pi-\theta$, depending on $\{\alpha,\beta\}$
(see Fig. \ref{fig:model}). Indeed, insertion of the above 
gaps into our model produces conductance spectra that display 
a ZBCP for \eg $\alpha=\beta=0$, as can be seen in Fig. \ref{fig:fig2}. In this case, as in Fig. \ref{fig:fig1}, 
the conductance spectra also allows for the magnitude of 
the superconducting gaps to be read out, although the features 
are not as clear as those seen in Fig. \ref{fig:fig1}.
\par
From the results of Figs. \ref{fig:fig1} and \ref{fig:fig2}, 
it is clear that the conductance spectra $G(E)$ exhibit strong dependence on the exchange energy, while the relative orientation of the gaps is insignificant for the superconducting gaps
Eq. (\ref{eq:gaps_1}). Thus, our results will shed light on the magnitude of the various components of the superconducting gaps and their relative orientations in $\vk$-space if the gaps display symmetry/antisymmetry under $\theta\to\pi-\theta$ for some orientation. Moreover, we are dealing with an easily 
observable effect, since distinguishing between the peaks 
occuring for various values of $R_\Delta$ requires a resolution 
of order ${\cal{O}}(10^{-1}\Delta_{\uparrow,0})$, which typically corresponds to $0.1-1$ meV. These structures should be then readily resolved with present-day STM technology. However, in order to do justice to the experimentalist, it should be pointed out that a challenge with respect to tunneling junctions is dealing with non-idealities at the interface which may affect the conductance spectrum. Also, the importance of spin-flip processes in the vicinity of the interface (if such are indeed present) has recently been pointed out \cite{kreuzer}.

\section{Conclusion}\label{sec:conclusion}
In summary, we have studied the conductance spectra 
$G(E)$ for systems consisting of a ferromagnetic metal 
separated from a non-unitary $p$-wave FMSC by a thin, 
insulating barrier. We have considered the cases of 
weak, intermediate, and strong barriers, and considered 
three different strengths of the 
ferromagnetic exchange energy. We have considered two 
classes of anisotropic spin-triplet superconducting gaps, 
with results given in Figs. (\ref{fig:fig1}) and (\ref{fig:fig2}). 
Our results show how the magnitude of the superconducting gaps $\Delta_{\sigma},\;\sigma=\uparrow,\downarrow$ may be inferred 
from the conductance spectra. Moreover, the class of superconducting  gaps given in Eq. (\ref{eq:gaps_1}) renders 
$G(E)$ insensitive to $\{\alpha, \beta \}$, the results are shown 
in Fig. (\ref{fig:fig1}) for $\{ \alpha, \beta \} =0$. 
Conversely, the orientations of the $p$-wave gaps modelled 
by Eq. (\ref{eq:gaps_2}), specific values of $\{\alpha,\beta\}$ 
are essential to the formation of ZBCPs in $G(E)$ in Fig. (\ref{fig:fig2}), and the characteristic 
behavior at $E=\Delta_{\sigma,0},\;\sigma=\uparrow,\downarrow$. These results should provide useful insights in determining both 
the relative orientation between the superconducting gaps, 
as well as their magnitude, in ferromagnetic spin-triplet superconductors.\\
\section*{Acknowledgments}
This work was
  supported by the Norwegian Research Council Grants No. 157798/432
  and No. 158547/431 (NANOMAT), and Grant No. 167498/V30 (STORFORSK).\\

\end{document}